# Emergence of diverse epidermal patterns via the integration of the Turing pattern model with the majority voting model

Short Title: Integration of the Turning pattern and majority voting models

Takeshi Ishida[1]

1 Department of Ocean Mechanical Engineering, National Fisheries University, Shimonoseki, Yamaguchi, Japan

2-7-1, Nagata-honmachi, Shimonoseki, Yamaguchi, 759-6595, Japan

*Corresponding author

Takeshi Ishida

E-mail: ishida@fish-u.ac.jp

# Abstract

The Turing pattern model is a type of reaction–diffusion (RD) model that can explain pattern formation based on the interaction between active and inhibitory factors. Pattern formation was first identified in an actual animal using the Turing pattern model in the 1990s in sea anemone. However, it remains unclear whether all epidermal patterns in animals can be explained by the Turing pattern model (or its related models). Even among fish, some species do not clearly follow Turing patterns. For example, the body patterns of the ornamental carp Nishiki goi produced in Japan vary randomly among individuals. Therefore, it is difficult to predict the pattern of the offspring based on the parent fish. Such a randomly formed pattern could be explained by a majority voting model. This model is a cellular automaton model that uses the sum rule of the Moore neighborhood based on the majority vote. Thus, the epidermal pattern of fish can be explained by either the Turing pattern or majority voting model. Nevertheless, the mechanism by which fish use these two different models remains unclear. Furthermore, patterns from these two types of models are detected among extremely closely related species. It is difficult to believe that completely different epidermal formation mechanisms are used among species of the same family. Therefore, there may be a more basic model that can produce patterns for both models. In this study, both models were represented by cellular automata, and their integration led to the proposal of a new model. By adjusting the parameters, this integrated model could create patterns that were equivalent to both the Turing pattern and majority voting models. After adjusting the intermediate parameter values of these two models, it was possible to create various patterns that were more diverse than those created by each model.

# Author Summary

Animal skin patterns are increasingly thought to be explained by the Turing pattern model proposed by Alan Turing. The Turing pattern model, which is a self-organizing model, can produce spotted or striped patterns. However, there are many animal patterns that do not correspond to the Turing pattern. For example, Holstein cattle have patterns that vary randomly from individual to individual. Such a pattern could be explained by a majority voting model. This model is a type of cellular automaton model that counts the surrounding states and transitions to the state with the higher number of states. In this study, the Turing pattern model and the majority voting model were represented by cellular automata, and then a new model integrating these two models was proposed. This integrated model is equivalent to both the Turing pattern model and the majority voting model by adjusting the parameters. Although this integrated model is extremely simple, it can produce a wider variety of patterns than either of the models.

# Introduction

The Turing pattern model is a type of reaction–diffusion (RD) model that was introduced by Turing in 1952 [1]. This model explains pattern formation based on the interaction between active and inhibitory factors. Two types of diffusion coefficient substances (morphogens) are assumed as these factors in the model. In the 1980s, Meinhardt [2] demonstrated that the Turing model can create various patterns via computer simulation. However, owing to the lack of concrete experimental evidence, it was not recognized for a long time as a model that could explain pattern formation in living organisms. Instead, Wolpert's "morphogen gradient model" [3] [4] for the morphogenesis of organisms was considered the dominant model. However, even Wolpert's model could not explain the robustness of the actual morphogenesis of organisms because of the dependence on initial values and vulnerability to disturbances. Thus, a definitive model for such morphogenesis remains unavailable.

Pattern formation was first identified in an actual animal using the Turing pattern model in the 1990s by Kondo and Asai [5] in sea anemone (*Pomacanthus imperator*). Regarding hybrids, Miyazawa et al. [6] compared the patterns of pure and hybrid species of salmonid fish and reported that each pattern could be explained by solving the Turing model equation and that the hybrid pattern could be reproduced by considering intermediate values of the parameters that reproduced the patterns of pure species.

Conversely, proteins or chemicals that are responsible for morphogenesis have not been identified, although such candidate substances (e.g., signaling factors such as TGF-β, Wnt, and Dkk [7, 8] and *Hox* gene products [9]) have been reported. Recent experimental studies have shown that the function of morphogens is not limited to the distribution of chemical concentrations, and they are involved in cell–cell interactions [10-13]. These experimental studies revealed that diffusion of chemicals is not the factor that forms Turing patterns; instead, these patterns include the autonomous movement of pigment cells [14] or cell-to-cell signaling via cell protrusions [15, 16]. Although the biochemical mechanisms of pattern formation remain unclear, experimental manipulation of patterns elucidated that the formation of certain body patterns is consistent with the models of the RD equation, such as the Turing pattern model; this is commonly accepted among biologists [17]. Furthermore, pattern formation is possible even in the absence of chemical diffusion if the conditions required for the interaction between local activation and long inhibition are satisfied [18].

Computer models have also been developed that reproduce more faithfully the realistic pattern formation processes that have been elucidated based on these experiments. For example, several simulations have been developed using agent-based models to reproduce zebrafish pigment pattern formation [19-22]. Vasilopoulos and Painter [23] also constructed a model with interacting cell protrusions and observed that, even if the protrusions are not anisotropic, the adjustment of their length and density can produce patterns that are similar to the RD model. In addition, Marcon et al. [24] investigated whether a stable stationary wave pattern can be generated in a three-factor RD system. In turn, Ishida et al. [25] developed the pufferfish skin patterns model using a cellular automata (CA) model. This CA model was based on Turing patterns through the exchange of binary values between neighboring cells. Despite the simplicity of the model, which uses five parameters (three parameters related to basic color pattern and two parameters for creating a large black spot), it was used to produce the skin patterns of *Takifugu*.

As a more generalized model, Kondo proposed the Kernel–Turing (KT) model [26], which uses distance and response profiles (i.e., kernels) to indicate activity and inhibition, and performs convolution integrals of these parameters to generate Turing patterns. Simulations of the KT model with kernels of various shapes showed that, in addition to being able to generate all standard patterns, i.e., stable 2D patterns (spots, stripes, and networks), it can also generate complex patterns that are difficult to generate using conventional Turing models.

Thus, models of the formation of animal epidermal patterns have long been studied based on the Turing model. Nevertheless, can we assume that all epidermal patterns in animals can be explained by the Turing pattern model (or similar relate models)? Previous studies have classified the body patterns of various fish [27], and it has been reported that the more typical Turing pattern is limited to a few of these species. The KT model [26] can also form derivative patterns other than the typical Turing pattern. However, even among fish, some individuals clearly do not exhibit Turing patterns, differing significantly from the patterns that can be generated using these models. For example, the body pattern of the ornamental carp Nishiki goi (*Cyprinus carpio*) produced in Japan varies randomly from individual to individual. Nishiki goi is the generic name of a variety of carp that has been improved for use as an ornamental fish. Various patterns have been created by crossbreeding these fish, including two-color red and white patterns, and three-color white, red, and black patterns, as shown in Fig 1. Patterns vary randomly from individual to individual, and it is difficult to predict the pattern of the offspring based on that of the parent fish. Although genetic studies of carp body coloration have been published, such as [28], no studies of models of reproduction of the patterns are available.

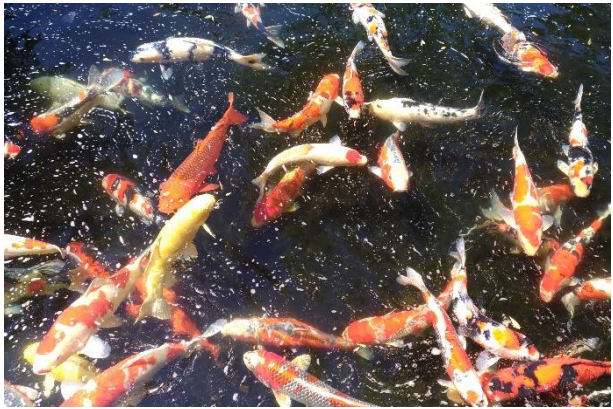

**Fig 1.** Examples of Nishiki goi patterns. Nishiki goi is an ornamental carp (*Cyprinus carpio*) in which various patterns have been created via crossbreeding, including two-color red and white patterns, and three-color white, red, and black patterns.

The example of the Nishiki goi patterns is similar to that of the black-and-white pattern of Holstein cows and the pattern of calico cats (tortoiseshell and white cats). The body patterns of these animals are genetically determined by the presence of two or three colors, whereas the shape of the pattern is randomly determined. In 2002, Qiu J. et al. [29] reported the birth of a somatic cell nuclear transplant clone of a cat: the nucleus was donated by a calico cat, the surrogate mother that gave birth to the cloned animal was a tiger cat, and the cloned cat was a calico cat. However, the patterns of the calico cat that donated the nucleus and that of the cloned cat exhibited different formations [29]. Although it is clear that genetic factors determine the number of colors in the pattern, the mechanism via which the pattern is formed at the body's surface remains under study.

The majority voting rule model proposed by Vichniac is one of an explanatory model in which these animal patterns are formed randomly [30]. The model is a cellular automaton model (the state can be 0 or 1) that uses the sum rule of the Moore neighborhood (8 adjacent cells) via majority vote, where the focus cell is 1 if the sum of the surrounding cells is 5 or more, and 0 if the sum is less than 5. The time evolution of this model from a random initial state strongly depends on the ratio or distribution of the 1 and 0 status in the initial state. It has been reported that, given proper initial placement, it is possible to form animal patterns using this model.

The epidermal pattern of Nishiki goi may also be explained by this majority rule model. The pattern differences observed among individuals are attributed to slight differences in the conditions present during the growth process, and can be thought of as corresponding to the sensitivity of the majority voting model to the initial values. From this, it can be inferred that the epidermal pattern of fish can be explained by either the Turing pattern model or the majority voting model. Nevertheless, how do fish use these two different models?

Patterns from these two types of models can also be found among extremely closely related species. For example, Nishiki goi is a member of the carp family; however, zebrafish, which is also a member of the carp family, has a Turing pattern; this pattern that has been used to investigate many morphological models. It is difficult to believe that completely different epidermal formation mechanisms are employed among species of the same family. Therefore, there may be a more basic model that can produce patterns for both models.

In this study, the Turing pattern and majority voting models were represented by CA, which led to the proposal of a new model that integrated these two models. After adjusting the parameters, this integrated model was equivalent to both previously mentioned models. Parameters that intermediate these two models can also be established. This integrated model produced a greater variety of patterns than either of the models. Although the model is extremely simple, it can produce a variety of patterns.

# Model

## Overview of the Turing pattern model

The Turing pattern model is a type of RD model that was introduced by Turing in 1952 [1], who considered morphogenesis as the interaction between activating and inhibiting factors. Typically, this model achieves self-organization through the different diffusion coefficients of two morphogens, which are equivalent to the activating and inhibiting factors. The general RD equations can be written as follows:

$$\frac{\partial u}{\partial t} = d_1 \nabla^2 u + f(u, v)$$

$$\frac{\partial v}{\partial t} = d_2 \nabla^2 v + g(u, v),$$

where u and v are the morphogen concentrations, functions f and g are the reaction kinetics, and $d_1$ and $d_2$ are the diffusion coefficients. Previous studies have considered various functions f and g; moreover, models such as the linear model, Gierer–Meinhardt model [31], and Gray–Scott model [32] have been used to produce typical Turing patterns.

# Representation of the Turing model using CA

In this study, instead of solving the RD differential equation directly, the CA model was used, which reproduces the Turing pattern with the characteristic "interactions between an activating factor and an inhibiting factor," which is a feature of the RD equation. CA models are discrete in both space and time. The state of the focal cell is determined by the states of the adjacent cells and transition rules. The advantage of CA models is that they can describe systems that cannot be modeled using differential equations.

Historically, various Turing-like CA patterns have been discovered. Markus [33] demonstrated that a CA model could produce the same output as did RD equations. The Young model [34] is one of the 2D totalistic models that bridges the RD equations and the CA model; this model is used to produce Turing patterns. Some other examples of the production of Turing patterns are provided below. Adamatzky [35] studied a binary-cell-state eight-cell neighborhood two-dimensional cellular automaton model with semitotalistic transitions rules. Dormann [36] also used a 2D outer-totalistic model with three states to produce a Turing-like pattern. In turn, Tsai [37] analyzed a self-replicating mechanism of Turing patterns using a minimal autocatalytic monomer–dimer system.

Young's CA model[34] uses a real number function, $v(r)$, to derive the distance effects, with two constant values within a grid cell: $u_1$ (positive) and $u_2$ (negative), as shown in Fig 2 (A). The $v(r)$ function is a continuous step function. The activation area, indicated by $u_1$, has a radius of $r_1$; and the inhibition area, indicated by $u_2$, has a radius of $r_2$ ($r_2 > r_1$) (Fig 2 (B)). The calculation begins by distributing black cells randomly on a rectangular grid. Subsequently, for each cell at position $R_0$ in 2D fields, the next cell state of $R_0$ is determined by the value of function $v(r)$. When $R_i$ is assumed to be a black cell within radius $r_2$ from the $R_0$ cell, and function $|R_0 - R_i|$ is assumed to be the distance between $R_0$ and $R_i$, the next cell state of $R_0$ is determined by the sum of the function $v(|R_0 − R_i|)$ value at all nearby black $R_i$ cells. If $\sum_i v(|R_0 − R_i|) > 0$, the grid cell at point $R_0$ is marked as a black cell; in turn, if $\sum_i v(|R_0 − R_i|) < 0$, the grid cell becomes a white cell. Finally, if $\sum_i v(|R_0 − R_i|) = 0$, the grid cell does not change state [34]. Young reported that a Turing pattern can be generated using these functions. Spot patterns or striped patterns can be created with relative changes between $u_1$ and $u_2$.

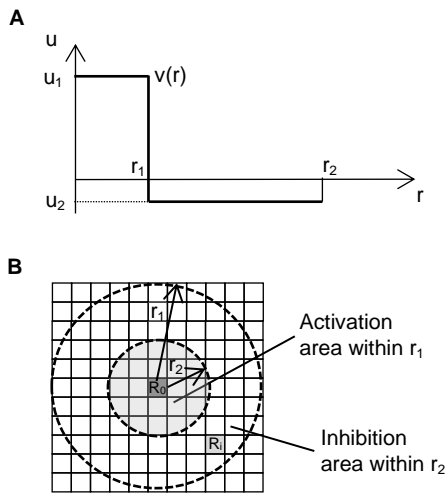

**Fig 2.** Outline of Young's model. (A) Function $v(r)$ is a continuous step function representing the activation area and the inhibition area. (B) The activation area has a radius indicated by $r_1$ and the inhibition area has an outer radius indicated by $r_2$.

In the present study, the method of Ishida was used [38], who converted the Young model into a simpler CA model. Ishida's method can be described as follows. In this Young model, let $u_1 = 1$ and $u_2 = w$ (here, $0 < w < 1$); furthermore, if the state of the cell is set to 0 (white) and 1 (black), this model can be arranged as follows. The state of cell i is expressed as $c_i(t)$ ($c_i(t) = [0, 1]$) at time $t$. The subsequent state at time $t + 1$, $c_i(t + 1)$, is determined by the states of the neighboring cells. Here, $N_1$ is the sum of the states of the domain within the $s_1$ meshes of the focal cell.

Similarly, $N_2$ is the sum of the states of the domain within the $s_2$ meshes of the focal cell, assuming that $s_1 < s_2$.

$$N_1 = \sum_{i=1}^{S_1} c_i(t)$$

$$N_2 = \sum_{i=1}^{S_2} c_i(t),$$

where $S_1$ and $S_2$ are the number of cells within the $s_1$ and $s_2$ meshes of the focal cell. In addition, $s_2 = 2s_1$ was assumed in this paper. Fig 3 provides a schematic representation of the total sum of states $N_1$ and $N_2$. The next state of the focal cell is determined by the following expression (1):

$$\text{Cell state at the next time step} = \begin{cases} 1: & \text{if } N_1 - N_2 \times w > 0 \\ \text{Unchange}: & \text{if } N_1 - N_2 \times w = 0 \quad (1), \\ 0: & \text{if } N_1 - N_2 \times w < 0 \end{cases}$$

where **w** and **s** are the two parameters that determine the Turing pattern.

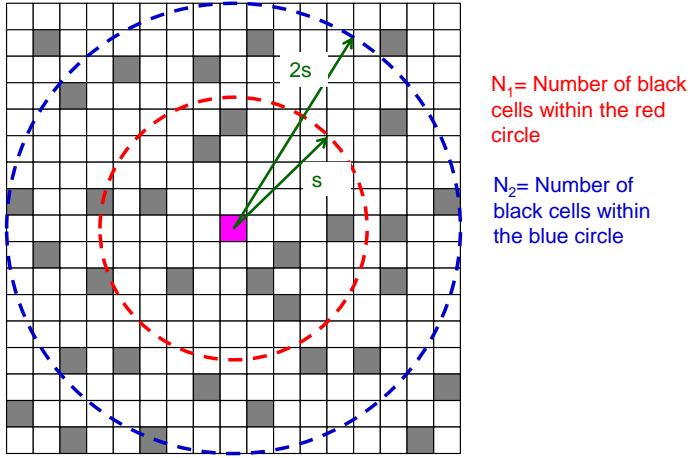

**Fig 3.** Schematic representation of the summing of states $N_1$ and $N_2$. Each grid cell has state 0 (white) or 1 (black). The inner area has a domain within the **s** grids of the focal cell and the outer area has a domain within the **2s** grids.

# Majority voting model using CA

Here, a model that applied the majority decision model proposed by Vichniac was considered [30]. The Vichniac model is a cellular automaton model that uses the sum rule of Moore neighborhoods (8 adjacent cells); however, in this model, cells in the $s_1$ range described in the previous section were considered, rather than cells that were merely adjacent. The focal cell value was set to 1 if the sum of the states in the cells in this $s_1$ range was greater than half the number of cells in the $s_1$ range, and to 0 if the sum was less than that. The results of the time evolution from a random initial state performed using this model strongly depended on the ratio or distribution of 1s and 0s in the initial state. When given the appropriate initial ratio, it is possible to form patterns that resemble animal body patterns. The equation for the majority voting model in this study was as indicated in (2).

$$\text{Cell state at the next time step} = \begin{cases} 1 & : \text{if } N_1 > \frac{N_0}{2} \\ \text{Unchange:} & \text{if } N_1 = \frac{N_0}{2} \\ 0 & : \text{if } N_1 < \frac{N_0}{2} \end{cases} \quad (2),$$

where $N_1$ is the sum of the states of the domain within the $s_1$ meshes of the focal cell and $N_0$ is the total number of cells within the $s_1$ range.

## Proposed integration model

After the integration of the two models presented in equations (1) and (2), they can be expressed using two parameters, **w** and **a**, by considering equations such as (3).

$$\text{Cell state at the next time step} = \begin{cases} 1 & : \text{if } N_1 - N_2 \times w > a \\ \text{Unchange:} & \text{if } N_1 - N_2 \times w = a \\ 0 & : \text{if } N_1 - N_2 \times w < a \end{cases} \quad (3)$$

In (3), when **w** > 0 and **a** = 0, the equation becomes the same as equation (1) and is a Turing pattern model. In turn, when **w** = 0 and **a** = $N_0/2$ in equation (3), the model is equivalent to equation (2) and is a majority rule model. It is also possible to construct a model that has intermediate parameters between the two models by varying **w** and **a**. This is the integrated model of the two models.

## Model with invariant regions at the boundary of the patterns

A variant of the majority voting model reported by Vichniac [30] has also been proposed. At the pattern boundary, where the majority decision is divided, the model is deformed so that the sum of the Moore neighborhoods is 0 when it should be 1 if the sum value is 5, and 1 when it should be 0 if the sum value is 4. This model works to weaken the effect of the majority decision at the pattern boundary. Based on this, a model was also devised that does not follow the Turing model or the majority rule model for the boundary domain of the pattern. Equation (4) is a model with a region that is invariant at the boundary of the pattern.

where a new parameter, **b**, has been added to set the range of the unchanged region.

$$\text{Cell state at the next time step} = \begin{cases} 1 & : \text{if } N_1 - N_2 \times w > \alpha\,(1+b) \\ 0 & : \text{if } N_1 - N_2 \times w < \alpha\,(1-b) \\ \text{Unchange} & : \text{otherwise} \end{cases} \quad (4)$$

## Calculation conditions

The model used 2D hexagonal grids (Fig 4) in which the application of transition rules was simple. Although square grids are generally used in 2D CA modeling, we also used hexagonal grids because they are isotropic compared with a square grid.

The models were implemented using the following conditions:

- Calculation field: 100 × 100 cells in hexagonal grids

- Periodic boundary condition

- Initial conditions: States 0 and 1 were placed randomly in each cell of the computational field with a probability of 0.5

- The range of $s_1$ was set to 3 cells from the focal cell, and the range of $s_2$ was set to 6 cells from the focal cell.

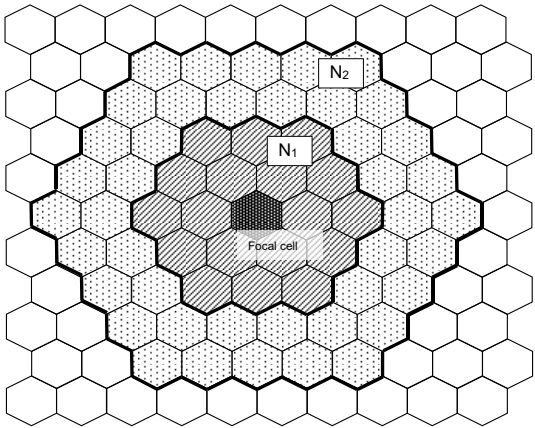

**Fig 4.** Hexagonal grid field. Here, $N_1$ is the domain within the $s_1$ meshes of the focal cell. Similarly, $N_2$ is the sum of the states of the domain within the $s_2$ meshes of the focal cell, assuming that $s_1 < s_2$.

# Results

## Parameter map

In the integrated model, after setting each parameter of **w** and **a**, the calculation was initiated from random initial values until the pattern became stationary. Figure 5 shows the parameter map, where the black cells indicate state 0 and blue cells indicate state 1.

The results of **w** = 0 and **a** = 1.0 at the top of the figure were equivalent to those of the majority voting model. In the case of the majority voting model with **w** = 0, the results changed sensitively according to the value of **a**. Therefore, the upper part of the whole map shows the results obtained when parameter a was altered slightly. When parameter **a** exceeded 1.0, the entire image was black (0), whereas when parameter **a** was <1.0, the entire image was blue (1).

The result of changing **w** with **a** = 0 in the rightmost column of the figure was equivalent to that of the Turing model. In regions where the **w** value was small, the cells were indicated in blue. As the w value increased, the cells appeared as black spots, changed to a striped pattern, became a blue-speckled pattern, and finally became fully black. These results are consistent with those of Ishida [38], and it is believed that the Turing pattern was reproduced.

The result of changing parameter **a** with each w value showed that the patterns were all black when the value of **a** was large. As the value of **a** decreased, the patterns became similar to those of the Turing pattern model with **a** = 0. It could also be observed that the effect of the majority voting model was stronger at approximately **w** = 0.1.

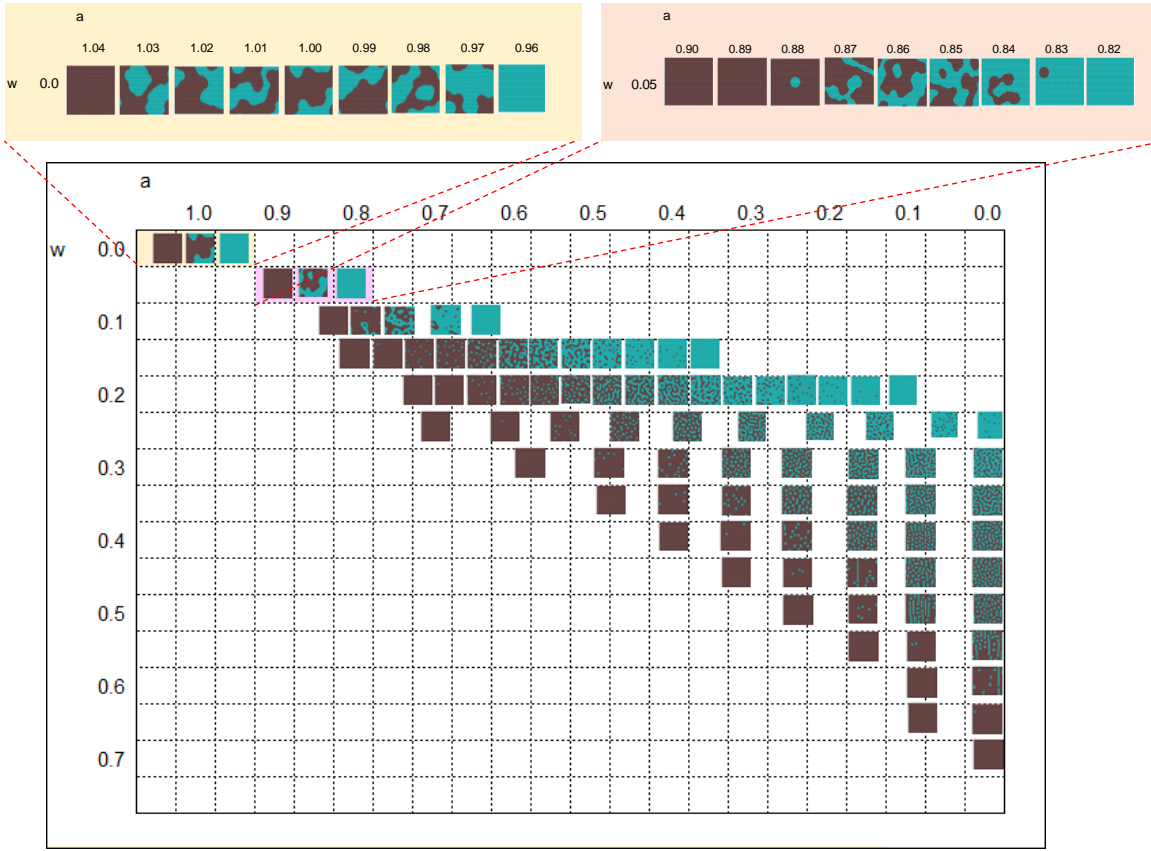

**Fig 5.** Parameter map with parameters **w** and **a**; the black cells indicate state 0 and the blue cells state 1. The results obtained for **w** = 0 and **a** = 1.0 indicated at the top of the figure were equivalent to those of the majority voting model. In the case of the majority voting model with **w** = 0, the results changed sensitively according to the value of **a**. Therefore, the upper part of the whole map presents the results obtained when **a** was varied slightly. The result obtained by changing w value with **a** = 0 in the rightmost column of the figure was equivalent to that of the Turing model.

## Initial value dependency

Smaller values of the parameter w (closer to the majority model) showed higher dependency on the initial value. Figure 6 presents the results of the initial value dependence in the equivalent model of majority voting with **w** = 0.0 and **a** = 1.0. These results were obtained when the cell states were placed randomly according to the specified black (0)/blue (1) ratio as the initial value. These are the results of five calculations at each ratio.

The results of this analysis revealed that a slight change in the black/blue ratio (from 0.47 to 0.53) as the initial value significantly changed the black and blue composition of the final pattern. Moreover, for the same black/blue ratio, the compositions of black and blue patterns in the final pattern tended to be similar, although the pattern changed for each calculation. The majority voting model with a black/blue ratio of 0.5 produced a pattern similar to the striped pattern.

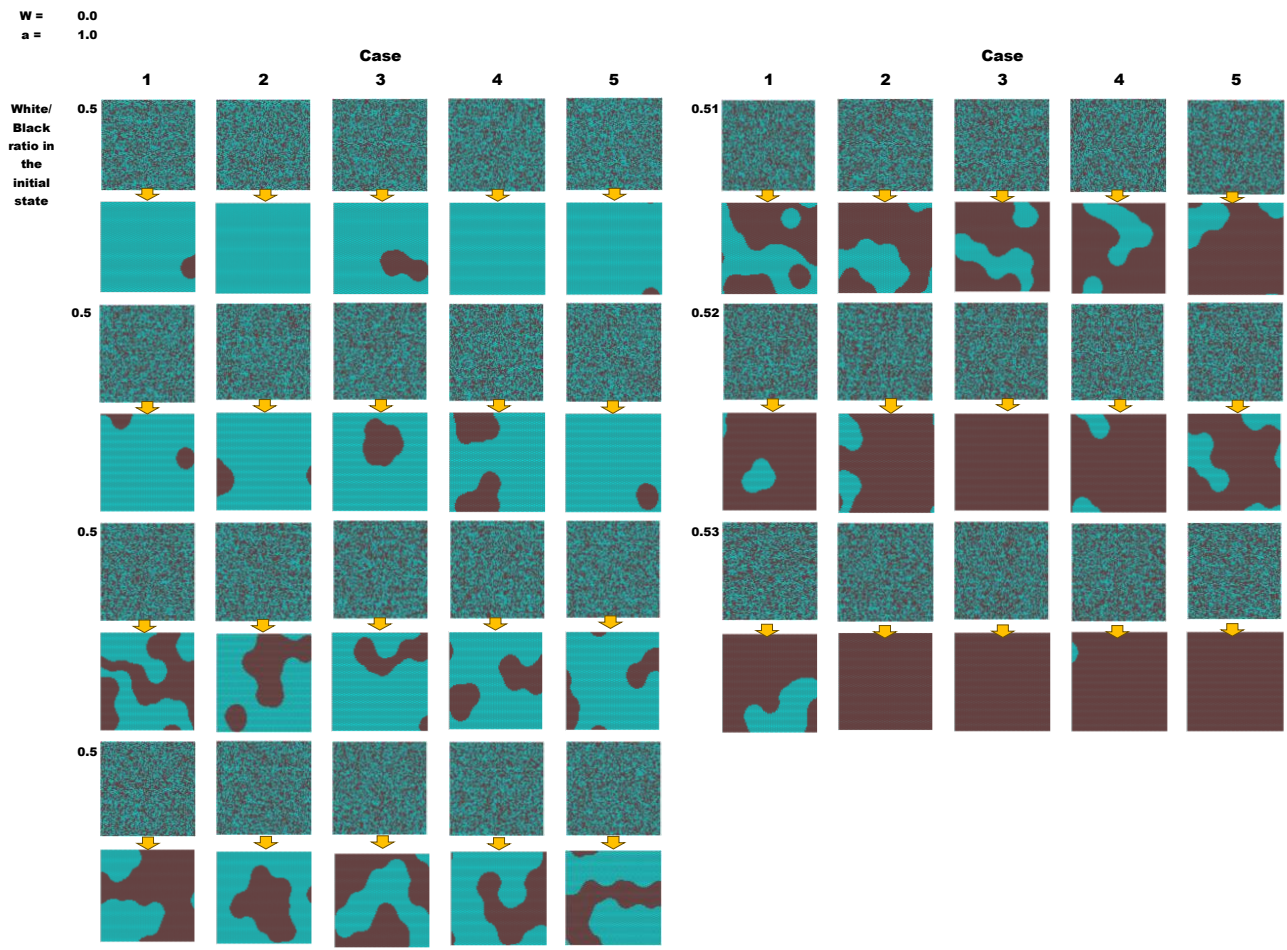

**Fig 6.** Dependence of the initial values in the majority voting model. These are the results of five calculations at each black (0)/blue (1) ratio as the initial value. The results indicated that a slight change in the black/blue ratio

(from 0.47 to 0.53) as the initial value can significantly change the black and blue composition of the final pattern.

# Results of the model with invariant regions at the boundary of the patterns

Figure 7 shows the results of the model that included regions in which the state was invariant at the boundaries of the patterns. The figure depicts the results obtained when **b** was altered under fixed conditions of the parameters

**w** and **a**. In this model, a larger value of parameter **b** yielded a larger region, in which the state that remained unchanged near the boundary of the pattern may expand. The results showed that the boundary of the pattern became more ambiguous as **b** increased.

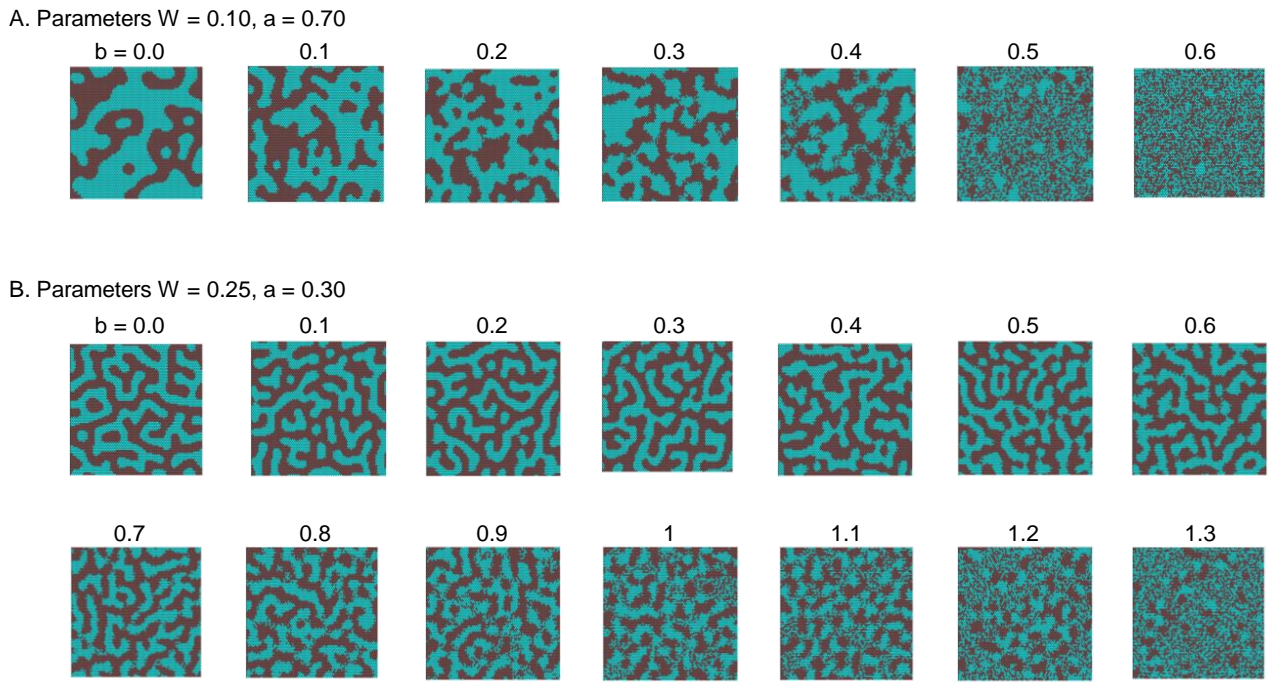

**Fig 7.** Results of the model with invariant regions at the boundary of the patterns; parameter **b** indicates the range of the unchanged region on the pattern's edge, with a larger parameter **b** yielding a larger region, in which the state that remained unchanged near the boundary of the pattern may expand. The results showed that the boundary of the pattern became more ambiguous as **b** increased.

## Discussion

In this integrated model, when the parameter **w** was 0, the model was assessed by the majority vote of the states of nearby cells. In contrast, when **w** was different from 0, the model tended to invert and incorporate the intensity of the information from distant locations (as indicated by the formula $N_1 - N_2 \times w$, where $N_2$ at distance distant location was negatively affected), thus creating a Turing pattern.

Figure 8A displays the results of the majority voting model. The fact that large patterns are formed at any initial value, despite being sensitive to the initial value, may explain the formation of Nishiki goi patterns. In general, a fixed number of colors are observed in Nishiki goi patterns; however, the appearance of the patterns varies greatly among individuals. This may explain why Nishiki goi patterns varied with slight changes in the conditions in epidermal cells during the growth process, corresponding to the initial value dependence of the model.

Figure 8B indicates the results obtained using parameters that were intermediate between those of the majority rule model and Turing pattern model, showing that patterns created using this approach could not be generated using the Turing model. The image on the right in Fig 8b provides an example of the pattern of the pufferfish *Takifugu vermicularis*, which is similar to the resulting pattern of the present model. Thus, a model with intermediate values of parameters could reproduce a wide variety of fish patterns.

Figure 8c provides an example of the calculation results based on a model in which the state was invariant in the boundary region of the patterns. We could create a pattern in which the boundaries of the pattern were intermittently connected, as observed in the case of the pufferfish *Takifugu stictonotus*.

Figure 8d depicts a speckled pattern with a typical Turing pattern. Under these conditions, it can be assumed that the pattern of the pufferfish *Takifugu alboplumbeus* was reproduced. In the Pufferfish family, the patterns vary greatly from species to species [6]; however, using this model, it was possible to create patterns in the same model by adjusting the parameters. Thus, the proposed integrated model can create a variety of patterns, from seemingly random patterns (such as those of Nishiki goi) to typical Turing model patterns, using a single model.

A. w = 0.0, a = 1.0    Nishiki goi

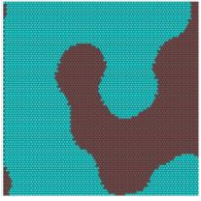 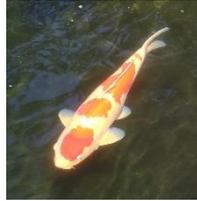 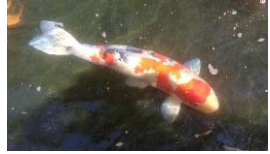 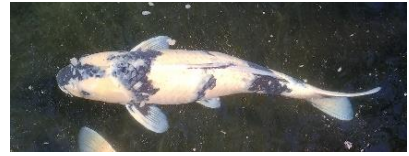

B. w = 0.1, a = 0.7    *Takifugu vermicularis*

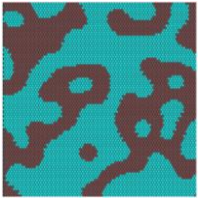 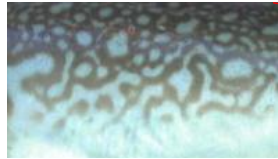 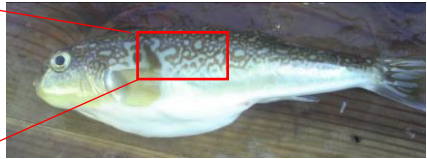

C. w = 0.25, a = 0.3, b = 1.1    *Takifugu stictonotus*

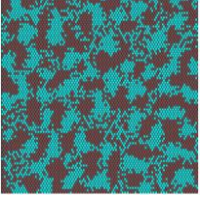 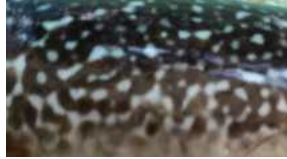 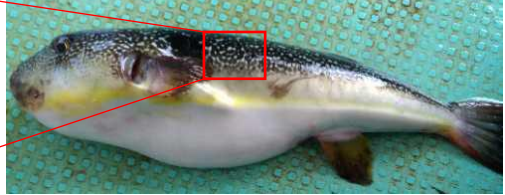

D. w = 0.50, a = 0.0    *Takifugu alboplumbeus*

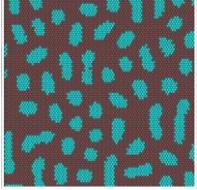 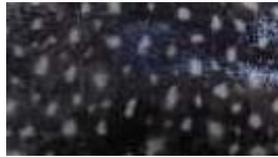 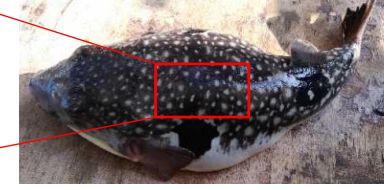

1 Web Fish Encyclopedia;   https://zukan.com/media/leaf/original/41224.jpg

**Fig 8.** Comparison of the calculation results of the integrated model with the actual patterns of the fish. It was found that the proposed integrated model created a variety of patterns, from seemingly random patterns (such as those of Nishiki goi) to typical Turing model patterns, using a single model.

In addition, the proposed model formed patterns by counting the states of the surrounding neighboring cells. It is a model that can be computed only via information transfer from neighboring cells. Therefore, it is possible to construct a model that corresponds to the recent experimental results of Turing patterns, such as mutual stimulation between cells. Although this model is simpler than the conventional computational models, it has the potential to create a wide variety of patterns.

Analysis of the overall parameter map (Fig 5) revealed that most of the map area was all black, i.e., an area that did not create patterns. This is consistent with the trend reported by Miyazawa [27], in which most fish had no pattern. Patterns are created exclusively when certain combinations of parameters are used. In particular, the patterns of Nishiki goi emerged only in an extremely narrow region of parameter a, under the condition of w = 0.0. Creating such a pattern is believed to be difficult with nature selection; these patterns can be explained consistently with the situation that it was found by artificial selection over a long period of time.

In this study, the Turing pattern and majority voting models were represented by a CA, which led to the proposal of a model that integrated these two models. By adjusting the parameters, this integrated model could create patterns that were equivalent to both the previously mentioned models. By setting the intermediate parameters values of the two models, it was possible to create a variety of patterns that were more diverse than those created by each model. Although this model is simpler than previously proposed models, it can create a variety of patterns. However, further research is warranted to determine whether this model is consistent with the mechanisms involved in the formation of fish patterns from a biological perspective.

In fact, many fish species dynamically change their patterns during the growth process from the juvenile to adult stages. A possible explanation for this finding is that cell–cell interactions on the epidermis change as the fish grows. It is believed that the application of this model may allow the examination of the large-scale changes in patterns associated with growth.

## Authors' contributions

The author, Ishida, conducted all the research.

## Competing interests

The author declares no conflict of interest regarding the publication of this paper.

## Funding Statement

This research was supported by grants from Japan Society for the Promotion of Science, KAKENHI Grant Number 19K04896.

## Acknowledgments

The authors would like to thank Enago (www.enago.jp) for the English language review.

## Availability of data and materials

The source code for the computational model of this study is available on GitHub at https://github.com/Takeshi-Ishida/Integrated-model-of-Turing-pattern-model-and-Majority-voting-model